\documentclass[twocolumn,tightenlines,showpacs,prd, floatfix,
preprintnumbers,amsmath,amssymb]{revtex4}

\usepackage{graphicx}

\usepackage{bm}

\begin{document}

\title{Gravitational wave bursts from cosmic (super)strings: 
Quantitative analysis and constraints}

\author{Xavier Siemens$^1$, Jolien Creighton$^1$, Irit Maor$^2$, 
Saikat Ray Majumder$^1$,
Kipp Cannon$^1$, Jocelyn Read$^1$}
\affiliation{
$^1$Center for Gravitation and Cosmology,
Department of Physics,
University of Wisconsin --- Milwaukee,
P.O. Box 413,
Wisconsin, 53201, USA\\
$^2$ CERCA, Department of Physics,
Case Western Reserve University, 10900 Euclid Ave,
Cleveland, OH 44106-7079
USA}

\date{\today}

\begin{abstract}
We discuss data analysis techniques that can be used in the search for
gravitational wave bursts from cosmic strings. When data from multiple
interferometers are available, we describe consistency checks that can
be used to greatly reduce the false alarm rates. We construct an
expression for the rate of bursts for arbitrary cosmic string loop
distributions and apply it to simple known solutions. The cosmology is
solved exactly and includes the effects of a late-time acceleration.
We find substantially lower burst rates than previous estimates
suggest and explain the disagreement. Initial LIGO is unlikely to
detect field theoretic cosmic strings with the usual loop sizes,
though it may detect cosmic superstrings as well as cosmic strings and
superstrings with non-standard loop sizes (which may be more
realistic). In the absence of a detection, we show how to set upper
limits based on the loudest event. Using Initial LIGO sensitivity
curves, we show that these upper limits may result in interesting
constraints on the parameter space of theories that lead to the
production of cosmic strings.
\end{abstract}


\maketitle

\section{Introduction}

Cosmic strings can form during phase transitions in the early universe
\cite{kibble76}. Although cosmic microwave background data has ruled
out cosmic strings as the primary source of density perturbations,
they are still potential candidates for the generation of a host of
interesting astrophysical phenomena including gravitational waves,
ultra high energy cosmic rays, and gamma ray bursts. For a review see
\cite{alexbook}.

Following formation, a string network in an expanding universe evolves
toward an attractor solution called the ``scaling regime'' (see
\cite{alexbook} and references therein). In this regime, the
statistical properties of the system, such as the correlation lengths
of long strings, scale with the cosmic time, and the energy density of
the string network becomes a small constant fraction of the radiation
or matter density.

The attractor solution arises due to reconnections, which for field
theoretic strings essentially always occur when two string segments
meet. Reconnections lead to the production of cosmic string loops,
which in turn shrink and decay by radiating gravitationally. This
takes energy out of the string network, converting it to gravitational
waves. If the density of strings in the network becomes large, then
strings will meet more often, thereby producing extra loops. The loops
then decay gravitationally, removing the surplus energy from the
network.  If, on the other hand, the density of strings becomes too
low, strings will not meet often enough to produce loops, and their
density will start to grow. Thus, the network is driven toward a
stable equilibrium.

Recently, it was realised that cosmic strings could be produced in
string-theory inspired inflation scenarios
\cite{tye1,tye2,alex15,tye3,cmp}. These strings have been dubbed
cosmic superstrings. The main differences between field theoretic
strings and superstrings are that the latter reconnect when they meet
with probabilities $p$ that can be less than $1$, and that more than
one kind of string can form. The values suggested for $p$ are in the
range $10^{-3}-1$ \cite{jjp}. The suppressed reconnection
probabilities arise from two separate effects. The first is that
fundamental strings interact probabilistically, and the second that
strings moving in more than $3$ dimensions may more readily avoid
reconnections: Even though two strings appear to meet in $3$
dimensions, they miss each other in the extra dimensions
\cite{alex15}. 

The effect of a small reconnection probability is to increase the time
it takes the network to reach equilibrium, and to increase the density
of strings $\rho$ at equilibrium.  The amount by which the density is
increased is the subject of debate, with predictions ranging from
$\rho \propto p^{-2}$, to $\rho \propto p^{-0.6}$
\cite{tye3,mairi,shellardlowp}. In this work we will assume the
density scales like $\rho \propto p^{-1}$, and that only one kind of
cosmic superstring is present \cite{DV2}.

Cosmic strings and superstrings can produce powerful bursts of
gravitational waves. These bursts may be detectable by first
generation interferometric earth-based gravitational wave detectors
such as LIGO and VIRGO \cite{DV2,DV0,DV1}. Thus, the exciting
possibility arises that a certain class of string theories may have
observable consequences in the near future \cite{pol1}. It should be
noted that if the hierarchy problem is solved by supersymmetry then
the detectability of bursts produced by cosmic superstrings is
strongly constrained due to dilaton emission \cite{babichev}.

The bursts we are most likely to be able to detect are produced at
cosmic string cusps. Cusps are regions of string that acquire
phenomenal Lorentz boosts, with gamma factors $\gamma \gg 1$. The
combination of the large mass per unit length of cosmic strings and
the Lorentz boost results in a powerful burst of gravitational
waves. The formation of cusps on cosmic string loops and long strings
is generic, and their gravitational waveforms are simple and robust
\cite{kenandi}. 

The three LIGO interferometers (IFOs) are operating at design
sensitivity. The VIRGO IFO is in the final stages of
commissioning. Thus, there arises a need for a quantitative and
observationally driven approach to the problem of bursts from cosmic
strings. This paper attempts to provide such a framework.

We consider the data analysis infrastructure that can be used in the
search for gravitational wave bursts from cosmic string cusps, as well
as cosmological burst rate predictions.

The optimal method to use in the search for signals of known form is
matched-filtering. In Sec. II we discuss the statistical properties of
the matched-filter output, template bank construction, an efficient
algorithm to compute the matched-filter, and, when data from multiple
interferometers is available, consistency checks that can be used to
greatly reduce the false alarm rate. In Sec. III we discuss the
application of the loudest event method to set upper limits on the
rate, the related issue of sensitivity, and background estimation. In
Sec. IV we derive an expression for the rate of bursts for arbitrary
cosmic string loop distributions, and apply it to the distribution
proposed in Refs. \cite{DV2,DV0,DV1}.  In Sec. V we compare our
estimates for the rate with results suggested by previous estimates
(in \cite{DV2,DV0,DV1}). We find substantially substantially lower
event rates.  The discrepancy arises primarily from our estimate of a
detectable burst amplitude, and the use of a cosmology that includes
the late-time acceleration. We also compute the rate for simple loop
distributions, using the results of Sec. IV and show how the
detectability of bursts is sensitive to the loop distribution.  In
Sec. VI we discuss how, in the absence of a detection, it is possible
to use the upper limits on the rate discussed in Sec. III to place
interesting constraints on the parameter space of theories that lead
to the production of cosmic strings. We show an example of these
constraints. We summarise our results and conclude in Sec. VII.

\section{Data analysis}

\subsection{The signal}

In the frequency domain, the waveforms for bursts of gravitational
radiation from cosmic string cusps are given by \cite{DV0},
\begin{equation}
h(f) = A |f|^{-4/3} \Theta(f_h-f)\,\Theta(f-f_l).
\label{eq:hf1}
\end{equation}
The low frequency cutoff of the gravitational wave signal, $f_l$, is
determined by the size of the feature that produces the
cusp. Typically this scale is cosmological.  In practice, however, the
low frequency cutoff of detectable radiation is determined by the low
frequency behaviour of the instrument (for the LIGO instruments, for
instance, by seismic noise). The high frequency cutoff, on the other
hand, depends on the angle $\theta$ between the line of sight and the
direction of the cusp. It is given by,
\begin{equation}
f_h \sim \frac{2}{\theta^3 L},
\label{eq:fh1}
\end{equation}
where $L$ is the size of the feature that produces the cusp, and in
principle can be arbitrarily large. The amplitude parameter of the
cusp waveform $A$ is,
\begin{equation}
A \sim \frac{G\mu L^{2/3}}{r},
\label{eq:A1}
\end{equation}
where $G$ is Newton's constant, $\mu$ is the mass per unit length of
strings, and $r$ is the distance between the cusp and the point of
observation. We have taken the speed of light $c=1$.

The time domain waveform can be computed by taking the inverse Fourier
transform of Eq.~(\ref{eq:hf1}). The integral has a solution in terms
of incomplete Gamma functions. For a cusp with peak arrival time at
$t=0$ it can be written as
\begin{eqnarray}
h(t) = 2 \pi A |t|^{1/3} &\{& i \Gamma(-1/3, 2\pi if_l t) + c.c.
\nonumber
\\
&-& [  i \Gamma(-1/3, 2\pi if_h t) + c.c.]\}.
\label{eq:ht1}
\end{eqnarray}
The duration of the burst is on the order of the inverse of
the the low frequency cutoff $f_l$, and the spike at $t=0$ is rounded
on a timescale $\sim 1/f_h$.

Eq.~(\ref{eq:ht1}) is useful to make a rough estimate the amplitude of
changes in the strain $h(t)$ that a cusp would produce. If we consider
a time near $t=0$, we can expand the incomplete Gamma
functions according to \cite{gradryz},
\begin{equation}
\Gamma(\alpha, x)=\Gamma(\alpha)+\sum_{n=0}^{\infty} \frac{(-1)^n
x^{\alpha+n}}{n!(\alpha+n)}.
\label{eq:gr8.354.2}
\end{equation}
Keeping the first term of the sum we see that the amplitude in
strain of the cusp is,
\begin{eqnarray}
\Delta h_{\text{cusp}} &=& 6 A (f_l^{-1/3} - f_h^{-1/3})
\nonumber
\\
&\approx& 6 A f_l^{-1/3},\,\,\,f_h \gg f_l.
\label{eq:cuspt5}
\end{eqnarray}

\subsection{Definitions and conventions}

The optimal method to use in the search for signals of known form is
matched-filtering. For our templates we take
\begin{equation}
t(f) = f^{-4/3} \Theta(f_h-f)\,\Theta(f-f_l),
\label{eq:templ1}
\end{equation}
so that a signal would have $h(f) = A\, t(f)$. We define the usual
detector-noise-weighted inner product \cite{cutlerflanagan},
\begin{eqnarray}
(x|y) \equiv 4 \Re \int_0^\infty df \,\frac{x(f)y^*(f)}{S_h(f)},
\label{eq:dorprod1}
\end{eqnarray}
where $S_h(f)$ is the single-sided spectral density defined by
$\langle n(f) n^*(f')\rangle = \frac{1}{2} \delta(f-f') S_h(f)$, where
$n(f)$ is the Fourier transform of the detector noise.

The templates of Eq.~(\ref{eq:templ1}) can be normalised by
the inner product of a template with itself,
\begin{equation}
\sigma^2=(t|t)
\label{eq:sigma1}
\end{equation}
so that,
\begin{equation}
{\hat t}=t/\sigma,\quad\mbox{and}\quad ({\hat t}|{\hat t})=1.
\label{eq:sigma2}
\end{equation}

If the output of the instrument is $s(t)$, the signal to noise ratio
(SNR) is defined as,
\begin{equation}
\rho \equiv (s|{\hat t}).
\label{eq:snr1}
\end{equation}
In general, the instrument output is a burst $h(t)$ plus some noise
$n(t)$, $s(t)=h(t)+n(t)$. When the signal is absent, $h(t)=0$, the SNR
is Gaussian distributed with zero mean and unit variance,
\begin{equation}
\langle \rho \rangle = \langle (n|{\hat t}) \rangle = 0, \,\,\,\,
\langle \rho^2 \rangle = \langle (n|{\hat t})^2 \rangle = 1.
\label{eq:snr2}
\end{equation}
When a signal is present, the average SNR is
\begin{eqnarray}
\langle \rho \rangle &=& \langle (h|{\hat t}) \rangle + 
\langle (n|{\hat t}) \rangle
= (A\sigma {\hat t}|{\hat t} ) = A\sigma,
\label{eq:snr7}
\end{eqnarray}
and the fluctuations have variance
\begin{eqnarray}
\langle \rho^2 \rangle - \langle \rho \rangle^2
= \langle (h+n|{\hat t})^2 \rangle - A^2\sigma^2=1.
\label{eq:snr8}
\end{eqnarray}
Thus, when a signal of amplitude $A$ is present, the signal to noise
ratio measured, $\tilde\rho$ , is a Gaussian random variable, with
mean $A\sigma$ and unit variance,
\begin{equation}
\tilde \rho =A\sigma \pm 1
\label{eq:snr10}
\end{equation}
The amplitude that we assign the event $\tilde A$ depends on the
template normalisation $\sigma$,
\begin{equation}
\tilde A = \frac{\tilde \rho}{\sigma}=A \pm \frac{1}{\sigma}.
\label{eq:amp1}
\end{equation}
This means that in the presence of Gaussian noise the relative
difference between measured and actual signal amplitudes will be
proportional to the inverse of the SNR,
\begin{equation}
\frac{\Delta A} {A}
= \pm \frac{1}{\langle \rho \rangle}.
\label{eq:Damp1}
\end{equation}
If an SNR threshold $\rho_{\text{th}}$ is chosen for the search, on
average only events with amplitude
\begin{equation}
A_{\text{th}} \ge \frac{\rho_{\text{th}}}{\sigma},
\label{eq:amp2}
\end{equation}
will be detected.

\subsection{Template bank}

Cosmic string cusp waveforms have two free parameters, the amplitude
and the high frequency cutoff. Since the amplitude is an overall
scale, the template need not match the signal amplitude. However, the
high frequency cutoff does affect the signal morphology, so a bank of
templates is needed to match possible signals.

The template bank, $\{t_i\}$, where $i=1,2,\dots N$, is one-dimensional,
and fully specified by a collection of high frequency cutoffs
$\{f_i\}$.  The number of templates $N$ should be large enough to
cover the parameter space finely. We choose the ordering of the
templates so that $f_i > f_{i+1}$.

Although in principle the high frequency cutoff can be infinite, in
practice the largest distinguishable high frequency cutoff is the
Nyquist frequency of the time-series containing the data,
$f_{\text{Nyq}}=1/(2\Delta t)$, where $\Delta t$ is the sampling time
of that time-series. The normalisation factor for the first template
is given by
\begin{equation}
\sigma^2_1=( t_1|t_1)
= 4 \Re \int_{f_l}^{f_{\text{Nyq}}} df \,\frac{f^{-8/3}}{S_h(f)}.
\label{eq:sigmatempl1}
\end{equation}
This template is the one with the largest $\sigma$, and thus the
largest possible SNR at fixed amplitude.

To determine the high frequency cutoffs of the remaining templates, we
proceed as follows. The fitting factor between two adjacent templates
$t_i$ and $t_{i+1}$, with high frequency cutoffs $f_i$ and $f_{i+1}$,
is defined as \cite{fittingfactor}
\begin{equation}
F=\frac{(t_i|t_{i+1})}
{\sqrt{( t_i| t_i)( t_{i+1}| t_{i+1})}}=1-\epsilon,
\label{eq:fftempl1}
\end{equation}
where $\epsilon$ is the maximum mismatch we choose to allow. The
maximum mismatch is twice the maximum fractional SNR loss due to
mismatch between a template in the bank and a signal, and should be
small. Since $t_{i+1}$ has a lower frequency cutoff than $t_{i}$,
\begin{equation}
( t_i| t_{i+1})= ( t_{i+1}| t_{i+1}).
\label{eq:sigmatempl2}
\end{equation}
Thus, the maximum mismatch is,
\begin{equation}
\epsilon = 1-\sqrt{\frac{(t_{i+1}| t_{i+1})}
{( t_i| t_i)}}=1-\frac{\sigma_{i+1}}{\sigma_i},
\label{eq:sigmatempl3}
\end{equation}
and once fixed, determines the high frequency cutoff $f_{i+1}$,
given $f_{i}$.  Thus the template bank can be constructed iteratively.

\subsection{Single interferometer data analysis}

A convenient way to apply the matched-filter is to use the {\it
inverse spectrum truncation} procedure \cite{invspectrunc}. The usual
implementation of the procedure involves the creation of an FIR
(Finite Impulse Response) digital filter for the inverse of the
single-sided power spectral density, $S_h(f)$, of some segment of
data. This filter, along with the template can then be efficiently
applied to the data via FFT (Fast Fourier Transform) convolution.
Here we describe a variation of this method in which the template as
well as $S_h(f)$ are incorporated into the digital filter.

Digital filters may be constructed for each template $j$ in the bank
as follows. First we define a quantity which corresponds to the square
root of our normalised template divided by the single-sided power
spectral density,
\begin{equation}
T_{j,1/2}(f) = \sqrt{ \frac{{\hat t}_j(f)} {S_h(f)}}.
\label{eq:sqrttemplf}
\end{equation}
This quantity is then inverse Fourier transformed to create,
\begin{equation}
T_{j,1/2}(t) = \int df e^{2\pi i f t} T_{j,1/2}(f).
\label{eq:sqrttemplt}
\end{equation}
At this point the duration of the filter is identical to the length of
data used to estimate the power spectrum $S_h(f)$. We then truncate
the filter, setting to zero values of the filter sufficiently far from
the peak. This is done to limit the amount of data corrupted by filter
initialisation. After truncation, however, the filter still needs to
be sufficiently long to adequately suppress line noise in the data
(such as power line harmonics).  The truncated filter is then forward
Fourier transformed, and squared, creating the frequency domain FIR
filter $T_j(f)$. The SNR is the time-series given by,
\begin{equation}
\rho_j (t) = 4 \Re \int_0^\infty df \, s(f) T_j(f) e^{2\pi i f t}.
\label{eq:rhot}
\end{equation}
where $s(f)$ is a Fourier transform of the detector strain data.

To generate a discrete set of events or triggers we can take the SNR
time series for each template $\rho_j (t)$ and search for clusters of
values above some threshold $\rho_{\text{th}}$. We identify these
clusters as triggers and determine the following properties: 1) The
SNR of the trigger which is the maximum SNR of the cluster; 2) the
amplitude of the trigger, which is the SNR of the trigger divided by
the template normalisation $\sigma_j$; 3) the peak time of the
trigger, which is the location in time of the maximum SNR; 4) the
start time of the trigger, which is the time of the first SNR value
above threshold; 5) the duration of the trigger, which includes all
values above the threshold; and 6) the high frequency cutoff of the
template. The final list of triggers for some segment of data can then
be created by choosing the trigger with the largest SNR when several
templates contain triggers at the same time (within the durations of
the triggers).

The rate of events we expect can be estimated for the case of white
Gaussian noise. In the absence of a signal the SNR is Gaussian
distributed with unit variance.  Therefore, at an SNR threshold of
$\rho_{\text{th}}$, the rate of events with $\rho>\rho_{\text{th}}$ we
expect with a single template is
\begin{equation}
R(>\rho_{\text{th}}) \approx \Delta t ^{-1}
{\text{Erfc}}(\rho_{\text{th}}/\sqrt{2})
\label{eq:rateofevents}
\end{equation}
where $\Delta t$ is the sampling time of the SNR time series, and Erfc
is the complementary error function. For example, if the time series
is sampled at $4096$~Hz and an SNR threshold of $4$ is used we expect
a false alarm rate of $R(>4) \approx 0.26$~Hz. The rate of
false alarms rises exponentially with lower thresholds.

\subsection{Multiple interferometer consistency checks}

When data from multiple interferometers is available, a coincidence
scheme can be used to greatly reduce the false alarm rate of events.
For example, in the case of the three LIGO interferometers, using
coincidence would result in a coincident false alarm rate of
\begin{equation}
R =  R_{\text{H1}} R_{\text{H2}} R_{\text{L1}} (2\Delta\tau_{\text{HH}}) 
(2\Delta\tau_{\text{HL}}),
\label{eq:coincidentrateofevents0}
\end{equation}
where $R_{\text{H1}}$ and $R_{\text{H2}}$ are the rate of false
triggers in the $4$~km and $2$~km IFOs in Hanford, WA, and
$R_{\text{L1}}$ is the rate of false triggers in the $4$~km IFO in
Livingston, LA. The two time windows $\Delta\tau_{\text{HH}}$, and
$\Delta\tau_{\text{HL}}$ are the maximum allowed time differences
between peak times for coincident events in the H1 and H2 IFOs, and
H1-L1 (or H2-L1), respectively. Eq.~(\ref{eq:coincidentrateofevents0})
follows trivially if the events are Poisson distributed. The factors
of $2$ multiplying the time windows arise from the fact that an event
in the trigger set of one instrument will survive the coincidence if
there is an event in the other instrument within a time $\Delta\tau$
on either side.  The time windows should allow for light-travel time
between sites, shifting of measured peak location due to noise, timing
errors and calibration errors. For example, if we take
$\Delta\tau_{\text{HL}} = 12$~ms and $\Delta\tau_{\text{HH}} = 2$~ms,
which allows for the light travel time of $10$~ms between the Hanford
and Livingston sites as well as $2$~ms for other uncertainties, and
the single IFO false alarm rate estimated in the previous subsection
($0.26$~Hz), we obtain a triple coincident false alarm rate,
\begin{equation}
R \approx  1.7 \times 10^{-6} \,\text{Hz},
\label{eq:coincidentrateofevents1}
\end{equation}
which is about $50$ false events per year.

Additionally, when two instruments are co-aligned, such as the H1 and
H2 LIGO IFOs, we can demand some degree of consistency in the
amplitudes recovered from the matched-filter (see, for example, the
distance cut used in \cite{amplitudecut}). If the instruments are not
co-aligned then the amplitudes recovered by the matched-filter could be
different. This is due to the fact that the antenna pattern of the
instrument depends on the source location in the sky. If the event is
due to a fluctuation of the noise or an instrumental glitch we do not
generally expect the amplitudes in two co-aligned IFOs to agree. The
effect of Gaussian noise to $1$ $\sigma$ on the estimate of the
amplitude can be read from Eq.~(\ref{eq:Damp1}). This equation can be
used to construct an amplitude veto. For example, to veto events in H1
we can demand
\begin{equation}
\frac{A_{\text{H1}}-A_{\text{H2}}} {A_{\text{H1}}}
< \pm \left(\frac{\delta}{\rho_{\text{H1}}} + \kappa \right),
\label{eq:Damp2}
\end{equation}
where $\delta$ is the number of standard deviations we allow and
$\kappa$ is an additional fractional difference in the amplitude that
accounts for other sources of uncertainty (such as the calibration).

These tests can be used to construct the final trigger set, a list of
survivors for each of our instruments which have passed all the
consistency checks.

\section{Upper limits, sensitivity and background}

In this section we describe the application of the loudest event
method \cite{loudestevent} to our problem, address the related issue
of sensitivity, and discuss background estimation.

We define the loudest event to be the survivor with the largest
amplitude. Suppose, in one of the instruments, we find the loudest
event to have an amplitude $A^*$. For an observation time $T$, on
average, the number of events we expect to show up with an amplitude
greater than $A^*$ is
\begin{equation}
N_{> A^*} = T \gamma, 
\label{eq:nevents}
\end{equation}
where the effective rate is defined as
\begin{equation}
\gamma = \int_0^{\infty} \epsilon (A^*,A)
\frac{dR(A)}{dA}dA.
\label{eq:nevents2}
\end{equation}
Here, $\epsilon (A^*,A)$ is the search efficiency, namely the fraction
of events with an optimally oriented amplitude $A$ found at the end of
the pipeline with an amplitude greater than $A^*$.  Aside from sky
location effects, the measured amplitude of events is changed as a
result of the noise (see Eq.~(\ref{eq:Damp1})). The efficiency can be
measured by adding simulated signals to detector noise, and then
searching for them.  The efficiency accounts for the properties of the
population, such as the distribution of sources in the sky and the
different high frequency cutoffs (as we will see, $dR/df \propto
f^{-5/3}$ for cusps), as well as possible inefficiencies of the
pipeline.  The quantity
\begin{equation}
\frac{dR(A)}{dA}dA
\label{eq:drda111}
\end{equation}
is the cosmological rate of events with optimally oriented amplitudes
between $A$ and $A+dA$ and will be the subject of the next Section.

If the population produces Poisson distributed events, the probability
that no events show up with a measured amplitude greater than $A^*$,
in an observation time $T$, is
\begin{equation}
P=e^{-T\gamma}.
\label{eq:Pnevents}
\end{equation}
The probability, $\eta$, that at least one event with amplitude greater
than $A^*$ shows up is $\eta=1-P$, so that if $\eta=0.9$, $90\%$ of
the time we would have expected to see an event with amplitude greater
than $A^*$.

From Eq.~(\ref{eq:Pnevents}) we see that,
\begin{equation}
\ln (1-\eta)=-T\gamma,
\label{eq:ln1}
\end{equation}
Setting $\eta=0.9$, say, we have
\begin{equation}
\gamma_{90\%} \approx \frac{2.303}{T}.
\label{eq:gamma2}
\end{equation}
We say that the value of $\gamma < \gamma_{90\%}$ with $90\%$
confidence, in the sense that if $\gamma = \gamma_{90\%}$, $90\%$ of
the time we would have observed at least one event with amplitude
greater than $A^*$. If we take a nominal observation time of one year,
$T=3.2\times 10^{7}$~s, our upper limit statement becomes,
\begin{equation}
\gamma < \gamma_{90\%} \approx 2.303 / \text{year} \approx 7.3
\times 10^{-8} \text{s}^{-1}.
\label{eq:gamma90}
\end{equation}
This bound on the effective rate can be used to constrain the
parameters of cosmic string models that enter through the cosmological
rate $dR/dA$ in Eq.~\ref{eq:nevents2}.

The efficiency $\epsilon (A^*,A)$ is the fraction of events with
optimally oriented amplitude $A$ which show up in our final trigger
set with an amplitude greater than $A^*$. If $A\gg A^*$ then we expect
$\epsilon (A^*,A) \approx 1$, and when $A \ll A^*$, $\epsilon (A^*,A)
\approx 0$. If $A_{50\%}$ is the amplitude at which half of the
signals are found with an amplitude greater than $A^*$, we can
approximate the sigmoid we expect for the efficiency curve by,
\begin{equation}
\epsilon (A^*,A) \approx \Theta (A-A_{50\%}),
\label{eq:epapprox}
\end{equation}
meaning we assume all events in the data with an amplitude
$A>A_{50\%}$ survive the pipeline and are found with an amplitude
greater than $A^*$. This means we can approximate the rate integral
$\gamma$ of Eq.~(\ref{eq:nevents2}) as,
\begin{equation}
\gamma \approx \int_{A_{50\%}}^{\infty} \frac{dR(A)}{dA}dA = R_{>A_{50\%}},
\label{eq:gamma3}
\end{equation}
so that if $R_{>A_{50\%}} > \gamma_{90\%}$ for a particular model of
cosmic strings, the model can be excluded (in the frequentist sense)
at the $90\%$ level.  We expect $A_{50\%}$ to be proportional to
$A^*$. It is at least a factor of $\sqrt{5}$ larger, due to the
averaging of source locations on the sky \cite{sqrt5factor}, and
potentially even larger than this due inefficiencies of the data
analysis pipeline. If there are no significant inefficiencies then we
expect $A_{50\%} \approx
\sqrt{5} A^*$.

At design sensitivity the form of the Initial LIGO noise curve may be
approximated by \cite{ligosens},
\begin{eqnarray}
S_h(f) &=& \left[ 1.09 \times 10^{-41}  \left( \frac{30\text{Hz}}{f}
\right)^{28} \right.
\nonumber
\\
&+&1.44 \times 10^{-45} \left( \frac{100\text{Hz}}{f} \right)^{4}
\nonumber
\\
&+&  \left. 1.28 \times 10^{-46} 
\left( 1+  \left( \frac{f}{90\text{Hz}}\right)^2 \right)\right]\text{s}.
\label{eq:LIGOIdesign}
\end{eqnarray}
The first term is the so-called seismic wall, the second term comes
from thermal noise in the suspension of the optics, and the third term
from photon shot noise. In the absence of a signal, this curve can be
used to estimate the amplitude of typical noise induced events.

As pointed out in \cite{DV2}, a reasonable operating point for the
pipeline involves an SNR threshold of $\rho_{\text{th}}=4$.  The value
of the amplitude is related to $\rho$, the SNR, and $\sigma$, the
template normalisation,
\begin{equation}
\sigma^2 = 4 \Re \int_0^\infty df \,\frac{t(f)^2}{S_h(f)}
= 4 \int_{f_l}^{f_h} df \,\frac{f^{-8/3}}{S_h(f)}
\label{eq:sigma10}
\end{equation}
by Eq.~(\ref{eq:amp2}). If the loudest surviving event has an SNR
close to the threshold, $\rho_* \approx 4$, a low frequency cutoff at
$f_l=40$~Hz (the seismic wall is such that it's not worth running the
matched-filter below $40$~Hz), and a typical high frequency cutoff
$f_h=150$~Hz then the amplitude of the loudest event will be $A_*
\approx 4 \times 10^{-21}$s$^{-1/3}$. So we expect $A_{50\%} \approx
9\times 10^{-21}$s$^{-1/3}$ for the Initial LIGO noise curve. 

The question of the sensitivity of the search is subtly different. In
this case we are interested in how many burst events we can detect at
all, not just those with amplitudes greater than $A^*$.  The rate of
events we expect to be able to detect is thus,
\begin{equation}
\gamma_s = \int_0^{\infty} \epsilon (0,A) \frac{dR(A)}{dA}dA.
\label{eq:gammasens}
\end{equation}
The efficiency includes events of all amplitudes and if the
cosmological rate $dR/dA$ favours low amplitude events, this rate
could be larger than that given by Eq.~(\ref{eq:nevents}). Furthermore
this rate should be compared with a rate of $1/T$, rather than, say,
the $90\%$ rate of Eq.~(\ref{eq:gamma2}). The net effect is that the
parameter space of detectability can be substantially larger than the
parameter space over which we can set upper limits. It should be noted
that the $0$ in the fraction of detected events $\epsilon (0,A)$ in
Eq.~\ref{eq:gammasens} means that we only demand that events with
optimally oriented amplitudes $A$ show up in the final trigger (not
that their amplitudes be larger than $A^*$). There is in fact a
minimum detectable amplitude of events, given by
$A_{\text{min}}=\rho_{\text{th}}/\sigma_{\text{max}}$, where
$\sigma_{\text{max}}$ is the largest template normalisation used in
the search.

In the remainder of the paper we will use an amplitude estimate of
$A_{50\%} = 10^{-20}$s$^{-1/3}$ for Initial LIGO upper limit and
sensitivity estimates. Advanced LIGO is expected to be somewhat more
than an order of magnitude more sensitive then Initial LIGO, so for
the Advanced LIGO rate estimates we will use $A_{50\%} =
10^{-21}$s$^{-1/3}$. It should be noted that the loudest event
resulting from a search on IFO data could be larger than this, for
instance if it was due to a real event or a surviving instrumental
glitch.

The rate of events that we expect in the absence of a signal, called
the background, can be estimated either analytically (via
Eq.~(\ref{eq:coincidentrateofevents0}), in which the single IFO rates
are measured), or by performing time-slides (see, for example
\cite{slides}). In the latter procedure single IFO triggers are
time-shifted relative to each other and the rate of accidental
coincidences is measured.  Care must be taken to ensure the
time-shifts are longer than the duration of real events.  If the rate
of events is unknown, then one can look for statistically significant
excesses in the foreground data set. In this case a statistically
significant excess could point to a detection. However, as we shall
see, there is enough freedom in models of string evolution to lead to
statistically insignificant excesses in {\it any}
background. Therefore it is prudent to carefully examine the largest
amplitude surviving triggers regardless of consistency with the
background.

\section{The rate of bursts}

In this section we will derive the expression for the cosmological
rate as a function of the amplitude, which is needed to evaluate the
upper limit and sensitivity integrals (Eqs.~(\ref{eq:nevents}) and
(\ref{eq:gammasens})). To make this account self-contained we will
re-derive a number of the results presented in \cite{DV2,DV0,DV1}.
The main differences from the previous approach are that we derive an
expression for a general cosmic string loop distribution that can be
used when such a distribution becomes known, we absorb some
uncertainties of the model, as well as factors of order $1$, into a
set of ignorance constants, and we consider a generic cosmology with a
view to including the effects of a late time acceleration.

Numerical simulations show that at any cosmic time $t$, the network
consists of a few long strings that stretch across the horizon and a
large collection of small loops. In early simulations \cite{earlysims}
the size of loops, as well as the substructure on the long strings,
was at the size of the simulation resolution. At the time of the early
simulations, the consensus reached was that the size of loops and
sub-structure on the long strings was at the size given by
gravitational back-reaction.  Very recent simulations
\cite{RSB,shellardrecentsim,VOV} suggest that the size of loops is
given by the large scale dynamics of the network, and is thus
un-related to the gravitational back-reaction scale. Further work is
necessary to establish which of the two possibilities is correct.

In either case, the size of loops can be characterised by a
probability distribution,
\begin{equation}
n(l,t)dl,
\label{eq:nl1}
\end{equation}
the number density of loops with sizes between $l$ and $l+dl$ at a
cosmic time $t$. This distribution is unknown, though analytic
solutions can be found in simple cases (see \cite{alexbook} \S 9.3.3).

The period of oscillation of a loop of length $l$ is $T=l/2$. If, on
average, loops have $c$ cusps per oscillation, the number of cusps
produced per unit space-time volume by loops with lengths in the
interval $dl$ at time $t$ is \footnote{Here we have assumed that the
number of cusps per oscillation does not depend on the length of the
loop. This will be true provided that loops have the same shape
(statistically), regardless of their size.} 
\begin{equation}
\nu(l,t)dl = \frac{2c}{l}n(l,t)dl.
\label{eq:nl2}
\end{equation}
We can write the cosmic time as a function of the redshift,
\begin{equation}
t = H^{-1}_0 \varphi_t (z),
\label{eq:cosmict1}
\end{equation}
where $H_0$ is current value of the Hubble parameter and $\varphi_t
(z)$ is a function of the redshift $z$ which depends on the
cosmology. In \cite{DV2,DV0,DV1}, an approximate interpolating
function for $\varphi_t (z)$ was used for a universe which contains
matter and radiation. For the moment we will leave it
unspecified. Later we will compute it numerically for a more realistic
cosmology which includes the late-time acceleration.  We can therefore
write the number of cusps per unit space-time volume produced by loops
with sizes between $l$ and $l+dl$ at a redshift $z$,
\begin{equation}
\nu(l,z) dl = \frac{2c}{l}n(l,z)dl.
\label{eq:nl3}
\end{equation}

Now we would like to write the length of loops in terms of the
amplitude of the event an optimally oriented cusp from such a loop
would produce at an earth-based detector.  The amplitude we expect
from a cusp produced at a redshift $z$, from a loop of length $l$ can
be read off Eq.~(4.10) of \cite{DV1}. It is
\begin{equation}
A  = g_1 \frac{G\mu l^{2/3} H_0} {(1+z)^{1/3}  \varphi_r (z)}.
\label{eq:amp11}
\end{equation}
Here $g_1$ is an ignorance constant that absorbs the uncertainty on
exactly how much of the length $l$ is involved in the production of
the cusp, as well as factors of $O(1)$ that have been dropped from the
calculation (see the derivation leading up to Eq.~(3.12) in
\cite{DV1}). If loops are not too wiggly, we expect this ignorance
constant to be of $O(1)$. We have expressed $r(z)$, the amplitude
distance \cite{DV1}, as an unspecified cosmology dependent function of
the redshift $\varphi_r (z)$,
\begin{equation}
r(z)=H^{-1}_0 \varphi_r (z).
\label{eq:rofZ}
\end{equation}
The amplitude distance is a factor of $(1+z)$ smaller than
the luminosity distance.  Since,
\begin{equation}
l(A,z)=\left( \frac{A \varphi_r (z)(1+z)^{1/3}}{g_1 G\mu H_0}
\right)^{3/2}, \,\,\mbox{and} \quad
\frac{dl}{dA}=\frac{3}{2A}l,
\label{eq:lofAnZ}
\end{equation}
we can write the number of cusps per unit space-time volume which
produce events with amplitudes between $A$ and $A+dA$ from a redshift
$z$ as,
\begin{equation}
\nu(A,z) dA = \frac{3c}{A}n(l(A,z),z)dA.
\label{eq:nl4}
\end{equation}

As discussed in \cite{DV0}, we can observe only a fraction of all the
cusps that occur in a Hubble volume. Suppose we look, via
matched-filtering, for signals of the form of Eq.~(\ref{eq:hf1}). The
highest frequency we observe $f_h$ is related to the angle $\theta$
that the line of sight makes to the direction of the cusp. If $f_*$ is
the lowest high frequency cutoff that our instrument can detect with
confidence, the maximum angle the line of sight and the direction of a
{\it detectable } cusp can subtend is,
\begin{equation}
\theta_m = (g_2 f_* l)^{-1/3}.
\label{eq:theta}
\end{equation}
The ignorance constant $g_2$ absorbs a factor of about $2.31$
\cite{DV1}, other factors of $O(1)$, as well as the fraction of the
loop length $l$ that actually contributes to the cusp (see the
derivation leading up to Eq.~(3.22) of \cite{DV1}).  Just like for our
first ignorance constant $g_1$, if loops are not too wiggly, we expect
$g_2$ to be of $O(1)$. For bursts originating at cosmological
distances, red-shifting of the high frequency cutoff must be taken
into account. So we write,
\begin{equation}
\theta_m (z,f_*,l)= [g_2 (1+z)f_* l]^{-1/3},
\label{eq:theta2}
\end{equation}

The waveform in Eq.~(\ref{eq:hf1}), was derived in the limit $\theta
\ll 1$, so that cusps with $\theta \gtrsim 1$ will not have the form
of Eq.~(\ref{eq:hf1}) (indeed, they are not bursts at all). Therefore
cusps with $\theta \gtrsim 1$ should not contribute to the rate. Thus,
the fraction of cusps that are observable, $\Delta$, is
\begin{equation}
\Delta(z,f_*,l) \approx \frac{\theta_m^2(z,f_*,l)}{4}
\Theta(1-\theta_m(z,f_*,l)).
\label{eq:Delta0}
\end{equation}
The first term on the right hand side is the beaming fraction for the
angle $\theta_m$. At fixed $l$, the beaming fraction is the fraction
of cusp events with high frequency cutoffs greater than $f_*$.  The
$\Theta$ function cuts off cusp events that have $\theta_m \geq 1$; it
ensures we only count cusps whose waveform is indeed given by
Eq.~(\ref{eq:hf1}).

The rate of cusp events we expect to see from a volume $dV(z)$ (the
proper volume in the redshift interval $dz$), in an amplitude interval
$dA$ is {\footnote{In \cite{DV2,DV0, DV1}, the cutoff was placed on
the strain of cusps (at a fixed rate) rather than their rate, as we do
here. This is un-important, the purpose of the theta function is to
ensure we do not overestimate the rate.}}
\begin{equation}
\frac {dR}{dV(z)dA} = (1+z)^{-1} \nu(A,z) \Delta(z,f_*,A),
\label{eq:rate1}
\end{equation}
where $\Delta(z,f_*,A)=\Delta(z,f_*,l(A,z))$. The factor of
$(1+z)^{-1}$ comes from the relation between the observed burst rate
and the cosmic time: Bursts coming from large redshifts are spaced
further apart in time. We write the proper volume element as,
\begin{equation}
dV(z) = H^{-3}_0  \varphi_V (z) dz,
\label{eq:dv1}
\end{equation}
where $\varphi_V (z)$ is a cosmology dependent function of the
redshift. Thus, for arbitrary loop distributions, we can write the
rate of events in the amplitude interval $dA$, needed to evaluate the
upper limit integral Eq.~(\ref{eq:nevents}), and sensitivity integral
Eq.~(\ref{eq:gammasens}) as,
\begin{equation}
\frac {dR}{dA} = H^{-3}_0 \int^{\infty}_0 dz\, \varphi_V (z)
(1+z)^{-1} \nu(A,z) \Delta(z,f_*,A).
\label{eq:dRdA1}
\end{equation}

Damour and Vilenkin \cite{DV2,DV0,DV1} took the loop distribution to
be
\begin{equation}
n(l,t) = (p \Gamma G \mu)^{-1} t^{-3} \delta(l-\alpha t),
\label{eq:nlDV}
\end{equation}
where $p$ is the reconnection probability, which can be smaller than
$1$ for cosmic superstrings. The constant $\Gamma$ is the ratio of the
power radiated into gravitational waves by loops to $G\mu^2$, and thus
related to the lifetime of loops. It is measured in numerical
simulations to be $\Gamma \sim 50$.

In this loop distribution all loops present at a cosmic time $t$, are
of the same size $\alpha t$.  The distribution is consistent with the
assumption usually made in the literature that the size of loops is
given by gravitational back-reaction, and that $\alpha \sim \Gamma G
\mu$.  Recently it was realised that damping of perturbations
propagating on strings due to gravitational wave emission is not as
efficient as previously thought \cite{KandIGW}. As a consequence, the
size of the small-scale structure is sensitive to the spectrum of
perturbations present on the strings, and can be much smaller than the
canonical value $\Gamma G \mu t$ \cite{trio}.

If the value of $\alpha$ is given by gravitational back-reaction we can
write it as
\begin{equation}
\alpha = \varepsilon \left( \Gamma G \mu \right)^n.
\label{eq:alpha1}
\end{equation}
We use two parameters $\varepsilon$, introduced in \cite{DV2}, as well
as $n$ that can be used to vary the size of loops.  The parameter $n$
arises naturally from gravitational back-reaction models and is
related to the power spectrum of perturbations on long strings
\cite{trio}. For example, if the spectrum is inversely proportional to
the mode number, then $n=3/2$. This is the spectrum of perturbations
we expect if, say, the shape of the string is dominated by the largest
kink \cite{trio}.

As we have mentioned, recent simulations suggest that loops are
produced at sizes unrelated to the gravitational back-reaction scale
\cite{RSB,shellardrecentsim,VOV}. If this is the case, then the loops
produced survive for a long time, and the form of the distribution can
be computed analytically in the matter and radiation eras using some
simple assumptions (see \cite{alexbook} \S 9.3.3).

In general the rate integral needs to be computed, presumably
numerically, through Eq.~(\ref{eq:dRdA1}). In the case of the more
simple loop distribution of Eq.~(\ref{eq:nlDV}), it is convenient to
proceed slightly differently. In this case all loops at a given
redshift are of the same size, so we can directly associate amplitudes
with redshifts.

We can write the rate of cusp events we expect to see from the
redshift interval $dz$, from loops of length in the interval $dl$ as
\begin{equation}
\frac {dR}{dzdl} = H^{-3}_0 \varphi_V(z) (1+z)^{-1} \nu(l,z) \Delta(z,f_*,l).
\label{eq:rate11}
\end{equation}
Using Eq.~(\ref{eq:nl3}), Eq.~(\ref{eq:nlDV}) with $t$ expressed in
terms of the redshift, and integrating over $l$ we find,
\begin{eqnarray}
\frac {dR}{dz} &=& H_0 \frac{c(g_2 f_* H^{-1}_0)^{-2/3}}{2\alpha^{5/3} p
\Gamma G \mu} \varphi^{-14/3}_t(z) \varphi_V(z) (1+z)^{-5/3}
\nonumber
\\
&\times&  \Theta(1-\theta_m(z,f_*,\alpha H^{-1}_0 \varphi_t (z))
\label{eq:rate12}
\end{eqnarray}

At a redshift of $z$, all cusps produce bursts of the same amplitude,
given by replacing $l$ with $\alpha H^{-1}_0 \varphi_t (z)$ in
Eq.~(\ref{eq:amp11}).  Therefore the solution of,
\begin{equation}
\frac{\varphi^{2/3}_t (z)}{(1+z)^{1/3} \varphi_r (z)} = \frac{A H^{-1/3}_0
}{g_1 G\mu \alpha^{2/3}}
\label{eq:zofA1}
\end{equation}
for $z$ gives the redshift from which a burst of amplitude $A$
originates. This means we can perform the rate integral, say
Eq.~(\ref{eq:gamma3}), over redshifts rather than over amplitudes,
\begin{equation}
\gamma \approx \int^{z_{50\%}}_0 \frac{dR}{dz}dz,
\label{eq:gamma4}
\end{equation}
where $z_{50\%}$ is the solution of Eq.~(\ref{eq:zofA1}) for $A=A_{50\%}$.

To compute the cosmological functions, $\varphi_t(z)$, $\varphi_r
(z)$, and $\varphi_V (z)$, we use the set of cosmological parameters
in \cite{stdcosmoref}, which provide a good fit to recent cosmological
data.  The precise values of the parameters are not critical, but we
include them here for clarity.  They are, $H_0=73$~km~s$^{-1}$~Mpc
$^{-1}=2.4\times 10^{-18}$~s$^{-1}$, $\Omega_m=0.25$, $\Omega_r= 4.6 \times
10^{-5}$, and $\Omega_{\Lambda}=1-\Omega_m-\Omega_r$. The derivation
of the three cosmological functions is included in Appendix A for
completeness and result in the cosmological functions
Eqs.~(\ref{eq:phitexact}), (\ref{eq:phiAexact}), and
(\ref{eq:phiVexact}). In the remainder of the paper we will refer to
this cosmological model as the $\Lambda$ universe.

\section{Results I: Detectability}

\subsection{Comparison with previous estimates}

In \cite{DV2,DV0,DV1} Damour and Vilenkin considered a universe with
matter and radiation. They introduced a set of approximate
interpolating functions for $\varphi_t(z)$, $\varphi_r(z)$, and
$\varphi_V(z)$. They were
\begin{equation}
\varphi_t (z) = t_0 H_0 (1+z)^{-3/2} (1+z/z_{eq})^{-1/2},
\label{eq:phit1}
\end{equation}
\begin{equation}
\varphi_r (z) = t_0 H_0 \frac{z}{1+z},
\label{eq:phiA1}
\end{equation}
and \cite{AVPrivate},
\begin{equation}
\varphi_V (z) = (t_0 H_0)^3 10^2  z^2 (1+z)^{-13/2} (1+z/z_{eq})^{-1/2}.
\label{eq:phiV1}
\end{equation}
Damour and Vilenkin used $z_{eq} \approx 10^{3.9}$ as the
redshift of radiation matter equality and set the age of the universe
$t_0=10^{17.5}$~s. Notice that if we substitute Eqs.~(\ref{eq:phit1})
and (\ref{eq:phiV1}) into Eq.~(\ref{eq:rate12}) and set $\alpha=\Gamma
G \mu$ and $g_2=p=1$ we recover Eq.~(5.17) of \cite{DV1} (aside from a
factor of $z$ related to their use of a logarithmic derivative).

We will use Eqs.~(\ref{eq:phit1}), (\ref{eq:phiA1}) and
(\ref{eq:phiV1}) to compare the previous results \cite{DV2,DV0,DV1}
with those of a cosmology solved exactly including the effects of a
late time acceleration, i.e. the cosmological functions
Eqs.~(\ref{eq:phitexact}), (\ref{eq:phiAexact}), and
(\ref{eq:phiVexact}). Damour and Vilenkin considered the amplitude (in
strain) of events at a fixed rate of $1$ per year and compared that to
an SNR $1$, optimally oriented event (see, for example, the dashed
horizontal lines of Fig. 1 in \cite{DV1}). Instead, we have started by
estimating a detectable signal amplitude, and computed the event rate
at and above that amplitude.

Bursts from cosmic string cusps are Poisson distributed.  The quantity
we have computed, $\gamma$, is the rate of events in our data set with
amplitudes greater than $A_{50\%}$. In an observation time $T$ the
probability of not having such an event is $\exp(-\gamma T)$. Hence,
the odds of having at least least one event in our data set with
amplitude larger than our minimum detectable amplitude is,
\begin{eqnarray}
\eta=1-e^{-\gamma T}.
\label{detectionprob}
\end{eqnarray}

\begin{figure}[t]
{\centering
\resizebox*{1\columnwidth}{.45\textheight}
{\includegraphics{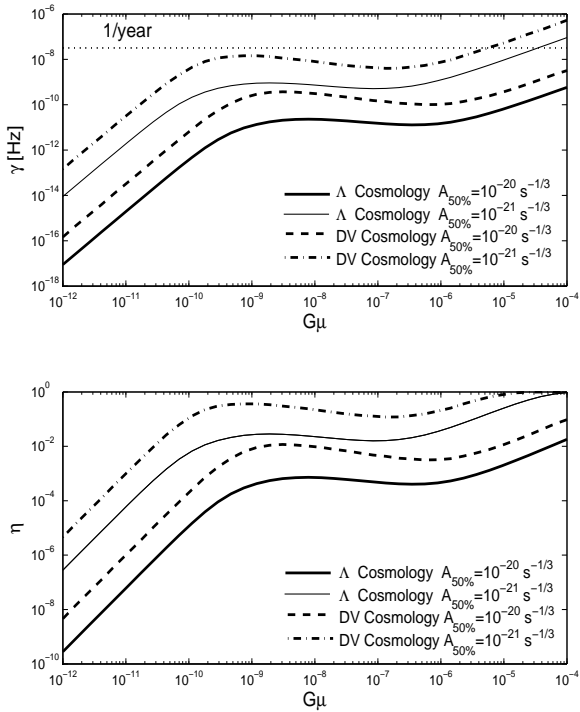}} \par}
\caption{ Plot of the rate of gravitational wave bursts, $\gamma$ (top
panel), and the probability $\eta$ of having at least least one event
in our data set with amplitude larger than $A_{50\%}$ in a year of
observation (bottom panel), as a function of $G\mu$. For all curves we
have set $\alpha=\Gamma G \mu$, $\Gamma=50$, $f_*=75$~Hz, $c=p=1$, and
the ignorance constants $g_1=g_2=1$.  The dash-dot and dashed curves
show $\gamma$ and $\eta$ computed with the Damour-Vilenkin
cosmological functions Eqs.~(\ref{eq:phit1}), (\ref{eq:phiA1}) and
(\ref{eq:phiV1}), with $A_{50\%} = 10^{-21}$~s$^{-1/3}$, and $A_{50\%}
= 10^{-20}$~s$^{-1/3}$ respectively. The thick and thin solid curves
show $\gamma$ and $\eta$ computed in a universe with a cosmological
constant with amplitudes $A_{50\%} = 10^{-20}$~s$^{-1/3}$, and
$A_{50\%} = 10^{-21}$~s$^{-1/3}$ respectively.  }
\label{fig:f1}
\end{figure}

Figure~\ref{fig:f1} shows the rate of burst events, $\gamma$, as well
as the probability $\eta$ of having at least least one event in our
data set with amplitude larger than $A_{50\%}$ for a year of
observation, as a function of $G\mu$ for two different models.  For
all curves we have set $\alpha=\Gamma G \mu$, $\Gamma=50$,
$f_*=75$~Hz, $c=p=1$, and the ignorance constants $g_1=g_2=1$. We will
refer to string models with these parameters as ``classic'', which is
appropriate for field theoretic strings with loops of size $l = \Gamma
G \mu t$. The dashed-dot and dashed curves of Fig.~\ref{fig:f1} show
$\gamma$ and $\eta$ computed using the Damour-Vilenkin cosmological
functions namely, Eqs.~(\ref{eq:phit1}), (\ref{eq:phiA1}) and
(\ref{eq:phiV1}).  For the dashed-dot curves we have used an amplitude
estimate of $A_{50\%} = 10^{-21}$~s$^{-1/3}$. This amplitude estimate
can be obtained using the Initial LIGO sensitivity curve, setting the
SNR threshold to $1$, and assuming all cusp events are optimally
oriented (as used for the dashed horizontal lines of Fig. 1 in
\cite{DV1}). This is also our estimate for the amplitude in the case
of Advanced LIGO. The dashed curves show $\gamma$ and $\eta$ computed
with the amplitude $A_{50\%} = 10^{-20}$~s$^{-1/3}$, which we feel is
more appropriate for Initial LIGO.  The thick and thin solid curves
show $\gamma$ and $\eta$ computed by evaluating the cosmological
functions (Eqs.~(\ref{eq:phitexact}), (\ref{eq:phiAexact}) and
(\ref{eq:phiVexact})) numerically for the $\Lambda$ universe (see
Appendix A). The thick solid curves correspond to our amplitude
estimate for Initial LIGO, and the thin solid curves to our estimate
for Advanced LIGO.

The functional dependence of the rate of gravitational wave bursts on
$G\mu$ is discussed in detail in Appendix B. Here we summarise those
findings. From left to right, the first steep rise in the rates as a
function of $G\mu$ of the dashed and dashed-dot curves of
Fig.~\ref{fig:f1} comes from events produced at small redshifts
($z \ll 1$). The peak and subsequent decrease in the rate starting
around $G\mu \sim 10^{-9}$ comes from events produced at larger
redshifts but still in the matter era ($1 \ll z \ll z_{eq}$). The
final rise comes from events produced in the radiation era ($z \gg
z_{eq}$).

For ``classic'' cosmic strings ($p=\varepsilon=n=1$), the matter era
maximum in our estimate for the rate of events at Initial LIGO
sensitivity is about $7 \times 10^{-4}$ events per year, which is
substantially lower than the rate $\sim 1$ per year suggested by the
results of Damour and Vilenkin \cite{DV2,DV0,DV1}. The bulk of the
difference arises from our estimate of a detectable amplitude. This is
illustrated by the dashed-dot and dashed curves of Fig.~\ref{fig:f1},
which use the same cosmological functions, and two estimates for the
amplitude, $A_{50\%} = 10^{-21}$~s$^{-1/3}$ and $A_{50\%} =
10^{-20}$~s$^{-1/3}$ respectively. Our amplitude estimate results in a
decrease in the burst rate by about a factor of $100$ at the matter
era peak. A more detailed discussion of the effect of the amplitude on
the rate can be found in Appendix B. The remaining discrepancy arises
from differences in the cosmology, as well as factors of $O(1)$ that
were dropped in the previous estimates, which account for a further
decrease by factor of about $10$. This is illustrated by the
difference between the dashed-dot and thin solid curves of
Fig.~\ref{fig:f1}, which use the same amplitude estimate $A_{50\%} =
10^{-21}$~s$^{-1/3}$, and the Damour-Vilenkin cosmological
interpolating functions (Eqs.~(\ref{eq:phit1}), (\ref{eq:phiA1}) and
(\ref{eq:phiV1})) and the $\Lambda$ universe functions
(Eqs.~(\ref{eq:phitexact}), (\ref{eq:phiAexact}) and
(\ref{eq:phiVexact})) respectively. When $z<<1$, the effects of a
cosmological constant are un-important and differences arise from
factors of $O(1)$ that were dropped in the previous estimates.  For $z
\gtrsim 1$, the differences arise from a combination of the effects of
a cosmological constant as well as factors of $O(1)$. The net effect
is that the chances of seeing an event from ``classic'' strings using
Initial LIGO data have dropped from order unity to about $10^{-3}$ at
the matter era peak. This is illustrated by the difference between the
dashed-dot and thick solid curves of Fig.~\ref{fig:f1}. The dashed-dot
curves were computed using the amplitude estimate $A_{50\%} =
10^{-21}$~s$^{-1/3}$ and the Damour-Vilenkin interpolating
cosmological functions, whereas the thick solid curves use an
amplitude estimate of $A_{50\%} = 10^{-20}$~s$^{-1/3}$ in the
$\Lambda$ universe.

\begin{figure}[t]
{\centering
\resizebox*{1\columnwidth}{.45\textheight}
{\includegraphics{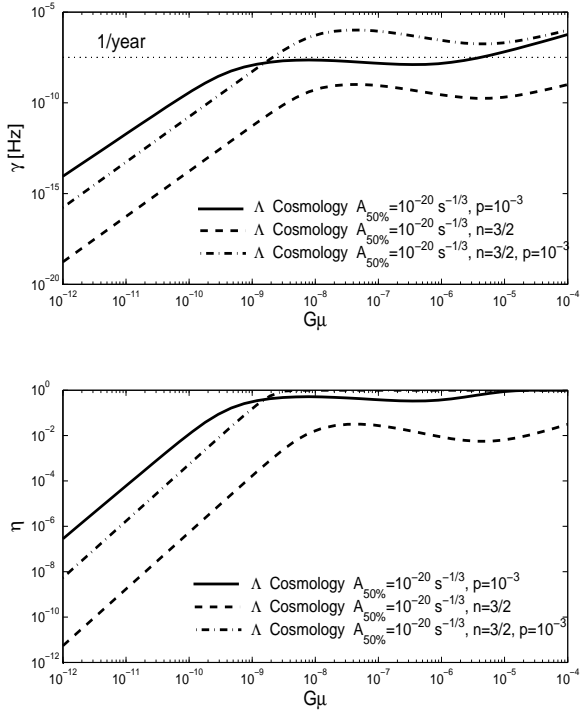}} \par}
\caption{Comparison of $\gamma$ (top panel) and $\eta$ with a year of
observation (bottom panel), as a function of $G\mu$ for several string
models. For all three curves an Initial LIGO amplitude estimate of
$A_{50\%} = 10^{-20}$~s$^{-1/3}$ has been used, and the cosmological
functions have been evaluated in the $\Lambda$ universe.  The model
parameters are identical to those of Fig.~\ref{fig:f1} except where
indicated.  The solid curves show $\gamma$ and $\eta$ computed with a
reconnection probability of $p=10^{-3}$.  The dashed curves show
$\gamma$ and $\eta$ computed with a size of loops given by
Eq.~(\ref{eq:alpha1}) with $\varepsilon=1$, and $n=3/2$. The
dashed-dot curves show the combined effect of a low reconnection
probability, $p=10^{-3}$, as well as a size of loops given by
Eq.~(\ref{eq:alpha1}) with $\varepsilon=1$, and $n=3/2$.}
\label{fig:f2}
\end{figure}

\begin{figure}[t]
{\centering
\resizebox*{1\columnwidth}{.45\textheight}
{\includegraphics{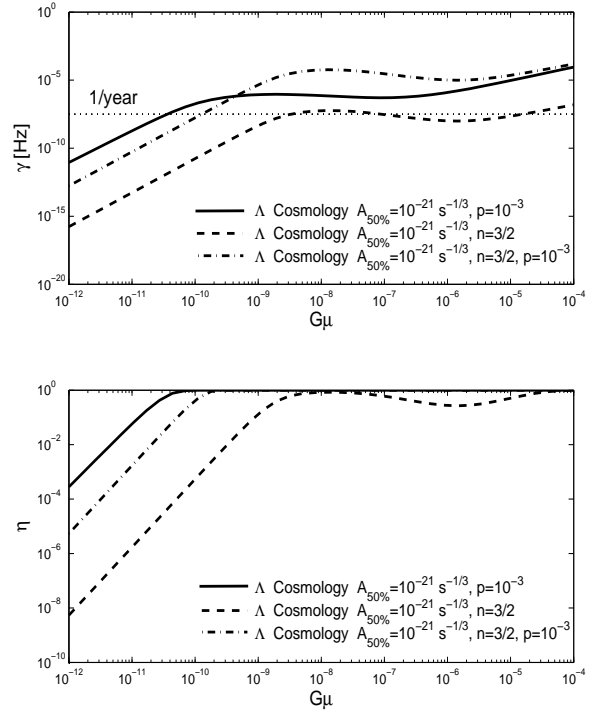}} \par}
\caption{
Same as Fig.~\ref{fig:f2} but with the Advanced LIGO amplitude estimate
of $A_{50\%} = 10^{-21}$~s$^{-1/3}$.}
\label{fig:f3}
\end{figure}

Cosmic superstrings, however, may still be detectable by Initial
LIGO. Furthermore, if the size of the small-scale structure is given
by gravitational back-reaction, reasonable estimates for what the size
of loops might be also lead to an enhanced rate of bursts.
Figure~\ref{fig:f2} illustrates this point. All curves use the Initial
LIGO amplitude estimate of $A_{50\%} = 10^{-20}$~s$^{-1/3}$, and the
cosmological functions Eqs.~(\ref{eq:phitexact}), (\ref{eq:phiAexact})
and (\ref{eq:phiVexact}) computed in the $\Lambda$ universe. The solid
curves show $\gamma$ and $\eta$ computed with a reconnection
probability of $p=10^{-3}$.  The dashed curves show $\gamma$ and
$\eta$ computed for loops with a size given by Eq.~(\ref{eq:alpha1})
with $\varepsilon=1$, and $n=3/2$. This is the value of $n$ we expect
when the spectrum of perturbations on long strings is inversely
proportional to the mode number, and the result we expect if the
spectrum of perturbations on long strings is dominated by the largest
kink \cite{trio}. The dashed-dot curves show the combined effect of a
low reconnection probability, $p=10^{-3}$, as well as a size of loops
given by Eq.~(\ref{eq:alpha1}) with $\varepsilon=1$, and $n=3/2$.  The
remaining parameters for all three curves are identical to those of
Fig.~\ref{fig:f1}. Advanced LIGO has a considerably larger chance of
making a detection of cosmic superstrings or field-theoretic strings
if loops are small. This is illustrated in Fig.~\ref{fig:f3} which
shows the same string models shown in Fig.~\ref{fig:f2}, with our
Advanced LIGO amplitude estimate of $A_{50\%} = 10^{-21}$~s$^{-1/3}$.

To summarise, we find the chances of detecting ``classic'' strings to
be significantly smaller than previous estimates suggest. Even
Advanced LIGO only has a few percent chance of detecting ``classic''
strings at the matter era peak (see the thin solid line around $G\mu
\sim 10^{-9}$ in Fig.~\ref{fig:f1}), though it has a good chance
of detecting cosmic superstrings and cosmic strings with small
loops as show in Fig.~\ref{fig:f3}. Initial LIGO requires the small
reconnection probability of cosmic superstrings and/or small loops to
attain a reasonable chance of detection.  It should be pointed out
that the ``classic'' string model may well be incorrect. If loop sizes
are given by gravitational back reaction then a reasonable guess for
the spectrum of perturbations on long strings leads to $n=3/2$. This
value, as shown on Fig.~\ref{fig:f2} has a substantially larger
chance of detection by Initial LIGO. The size of loops, however, may
not determined by gravitational back-reaction after all.

\subsection{Large loops}

Recent simulations \cite{RSB,shellardrecentsim,VOV} suggest that loops
are produced at sizes unrelated to the gravitational back-reaction
scale. One of the simulation groups \cite{RSB} finds a power law for
the loop distribution, while the other two groups
\cite{shellardrecentsim,VOV} find loops produced at a fixed fraction
of the horizon, with loop production functions peaking around
$\alpha\approx 10^{-3}$ \cite{shellardrecentsim}, and the
significantly larger $\alpha \approx 0.1$ \cite{VOV}. It should be
pointed out that the first two results \cite{RSB,shellardrecentsim}
are expanding universe simulations, whereas the results of the third
group \cite{VOV} come from simulations in Minkowski space.

Following formation, the length of loops shrinks due to gravitational
wave emission according to \cite{alexbook},
\begin{eqnarray}
l(t)=l_i - \Gamma G \mu (t-t_i),
\label{eq:loft}
\end{eqnarray}
where $l_i=\alpha t_i$, is the initial length, and $t_i$ is the time
of formation of the loop. The length goes to zero at time
\begin{eqnarray}
t_f = \left(\frac{\alpha}{\Gamma G \mu} +1\right) t_i.
\label{eq:tend}
\end{eqnarray}
Loops are long-lived when $t_f \gg t_i$, i.e. when $\alpha/(\Gamma G
\mu) \gg 1$. For $\alpha \approx 0.1$, using $\Gamma=50$ the lifetime 
of loops is long provided $G\mu \ll 2 \times 10^{-3}$, which covers
the entire range of astrophysically interesting values of $G\mu$. On
the other hand, if we take $\alpha \approx 5 \times 10^{-4}$, then
loops are long-lived only when $G\mu \ll 10^{-5}$. 

If the size of loops is given by gravitational back reaction, then
$\alpha$ is given by Eq.~(\ref{eq:alpha1}), and provided $n \ge 1$ all
loops are short-lived. This means we can use $n(l,t) \propto
\delta(l-\alpha t)$ (as we have so far) because the loop distribution
is dominated by the loops that just formed.

\begin{figure}[t]
{\centering
\resizebox*{1\columnwidth}{.45\textheight}
{\includegraphics{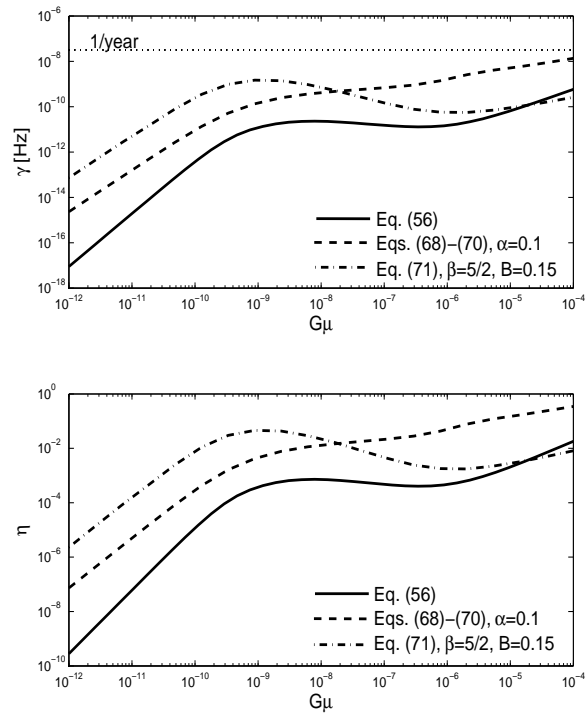}} \par}
\caption{ Plot of the rate of bursts $\gamma$ (top panel), and $\eta$,
the probability of having at least least one event in our data set
with amplitude larger than $A_{50\%}$ in a year of observation (bottom
panel), as a function of $G\mu$ for various loops distributions. For
all curves we have set $\Gamma=50$, $f_*=75$~Hz, $c=p=1$, and the
ignorance constants $g_1=g_2=1$.  As a reference we again show
$\gamma$ and $\eta$ computed using the loop distribution from
Eq.~(\ref{eq:nlDV}), according to Eq.~(\ref{eq:rate12}), in the
$\Lambda$-universe with $\alpha=\Gamma G \mu$ using the solid
curve. The remaining curves have been computed through
Eqs.~(\ref{eq:dRdA1}) and (\ref{eq:gamma3}) with $A_{50\%} =
10^{-20}$~s$^{-1/3}$.  The dashed curves show $\gamma$ and $\eta$
computed using the loop distribution of Eqs.~(\ref{eq:nltrad1}),
(\ref{eq:nltmat1}), and (\ref{eq:nltmat2}). The dashed-dot curves show
$\gamma$ and $\eta$ computed with Eq.~(\ref{eq:RSB}).}
\label{fig:f4}
\end{figure}

If loops are long-lived, the distribution can be calculated if a
scaling process is assumed (see \cite{alexbook}, \S 9.3.3 and \S
10.1.2). In the radiation era it is
\begin{eqnarray}
n(l,t)&=&\chi_r t^{-3/2} (l+\Gamma G \mu t)^{-5/2},  
\nonumber
\\
&\,& l < \alpha t,\,\,\, t<t_{\text{eq}}
\label{eq:nltrad1}
\end{eqnarray}
where $\chi_r \approx 0.4 \zeta \alpha^{1/2}$, and $\zeta$ is a
parameter related to the correlation length of the network found in
numerical simulations of radiation era evolution to be about $15$ (see
Table 10.1 in \cite{alexbook}).  The upper bound on the length arises
because no loops are formed with sizes larger than $\alpha t$.

In the matter era the distribution has two components, loops formed in
the matter era and survivors from the radiation era. Loops formed in
the matter era have lengths distributed according to,
\begin{eqnarray}
n_{1}(l,t)&=&\chi_m t^{-2} (l+\Gamma G \mu t)^{-2},
\nonumber
\\
&& \alpha t_{\text{eq}} 
- \Gamma G \mu (t-t_{\text{eq}}) < l < \alpha t,\,\,\, t>t_{\text{eq}}
\label{eq:nltmat1}
\end{eqnarray}
with $\chi_m \approx 0.12\zeta$, with $\zeta \approx 4$ (see Table
10.1 in \cite{alexbook}).  The lower bound on the length is due to the
fact that the smallest loops present in the matter era started with a
length $\alpha t_{eq}$ when they were formed and their lengths have
since decreased due to gravitational wave emission. Additionally there
are loops formed in the radiation era that survive into the matter
era. Their lengths are distributed according to,
\begin{eqnarray}
n_{2}(l,t)&=&\chi_r t_{\text{eq}}^{1/2} t^{-2} (l+\Gamma G \mu t)^{-5/2},
\nonumber
\\
&\,& l < \alpha t_{\text{eq}}- \Gamma G \mu  (t-t_{\text{eq}}),\,\,\, t>t_{\text{eq}},
\label{eq:nltmat2}
\end{eqnarray}
where the upper bound on the length comes from the fact that the
largest loops formed in the radiation era had a size $\alpha t_{eq}$
but have since shrunk due to gravitational wave emission.

The simulations in \cite{RSB} find a power law for the loop
distribution, $n(l) \propto l^{-\beta}$.  If all loops produced are
long-lived, meaning no loops are produced with sizes below $\Gamma G
\mu t$, the distribution we expect is
\begin{eqnarray}
n(l,t) = B t^{\beta-4} (l+\Gamma G \mu t)^{-\beta}.
\label{eq:RSB}
\end{eqnarray}
The most recent fits to their loop distribution find $\beta \approx
5/2$, in both the radiation and matter eras, and $B \approx 0.1, 0.2$
in the matter and radiation eras respectively \cite{Rprivate}.

\begin{figure}[t]
{\centering
\resizebox*{1\columnwidth}{.45\textheight}
{\includegraphics{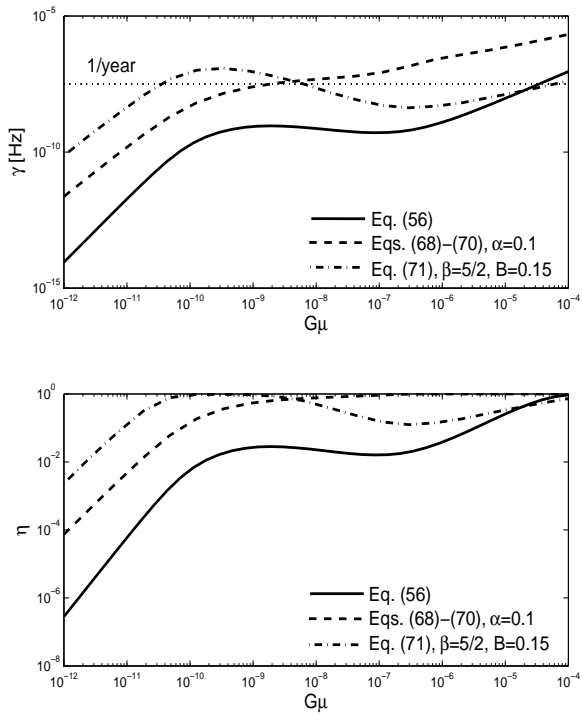}} \par}
\caption{
Same as Fig.~\ref{fig:f4} but using our Advanced LIGO amplitude
estimate $A_{50\%} = 10^{-21}$~s$^{-1/3}$.}
\label{fig:f5}
\end{figure}

We can use the results of Sec. IV to compute the rate of bursts for
these loop distributions. Figure~\ref{fig:f4} shows the burst rate
$\gamma$, and the probability $\eta$ of having at least least one
event in our data set with amplitude larger than $A_{50\%}$ in a year
of observation as a function of $G\mu$ for all the above
distributions. For all curves we have set $\Gamma=50$, $f_*=75$~Hz,
$c=p=1$, and the ignorance constants $g_1=g_2=1$.  We have evaluated
the cosmological functions in the $\Lambda$ universe.  As a reference,
we again show $\gamma$ and $\eta$ computed using the loop distribution
from Eq.~(\ref{eq:nlDV}), according to Eq.~(\ref{eq:rate12}), with
$\alpha=\Gamma G\mu$ using the solid curves. The remaining curves have
been computed through Eqs.~(\ref{eq:dRdA1}) and (\ref{eq:gamma3}) with
$A_{50\%} = 10^{-20}$~s$^{-1/3}$. The dashed curves show $\gamma$ and
$\eta$ computed using the loop distribution of
Eqs.~(\ref{eq:nltrad1}), (\ref{eq:nltmat1}), and (\ref{eq:nltmat2})
with $\alpha=0.1$. The dashed-dot curves show $\gamma$ and $\eta$
computed with Eq.~(\ref{eq:RSB}), where we have taken $\beta = 5/2$,
and $B=0.15$ as an approximation for both the radiation and matter
eras. Figure~\ref{fig:f5} shows the same loop distributions as
Fig.~\ref{fig:f4} but using our Advanced LIGO amplitude estimate 
$A_{50\%} = 10^{-21}$~s$^{-1/3}$.

The loop distributions shown here lead to a significant enhancement in
the rate. Note that we have not included the enhancement in the rate
of bursts for the case of cosmic superstrings $p<1$. Given the range
of results, it is important to determine whether the loop sizes at
formation are determined by gravitational back-reaction, or whether
they are large and we require a revised loop distribution.

\section{Results II: Constraints}

If no events can be positively identified in a search, we can use the
results of Sec. III to constrain the parameter space of theories
that lead to the production of cosmic strings. Unfortunately, as we
have mentioned, considerable uncertainties remain in models of cosmic
string evolution. Nevertheless, we can place constraints that are
correct in the context of a particular string model.  Here we will
illustrate the procedure for the loop distribution, Eq.~(\ref{eq:nlDV}),
and the resulting rate Eq.~(\ref{eq:rate12}).

We would like to absorb the ignorance constants $g_1$ and $g_2$ into
the parameters of the model. The two ignorance constants enter the
expression for the rate Eq.~(\ref{eq:rate12}) in three ways. First,
they affect the upper limit of the integral Eq.~(\ref{eq:gamma4})
through Eq.~(\ref{eq:zofA1}), which is,
\begin{equation}
\frac{\varphi^{2/3}_t (z)}{(1+z)^{1/3} \varphi_r (z)} = \frac{A H^{-1/3}_0
}{g_1 G\mu \alpha^{2/3}}.
\label{eq:zofA3}
\end{equation}
Secondly, they enter through $\theta_m$ in the theta function cutoff of the
rate,
\begin{equation}
\theta_m 
= (g_2 (1+z)f_* \alpha H_0^{-1} \varphi_t (z))^{-1/3},
\label{eq:theta3}
\end{equation}
and finally, the rate itself (Eq.~(\ref{eq:rate12})) is proportional to
$g_2^{-2/3}$.

\begin{figure}[t]
{\centering
\resizebox*{1\columnwidth}{.25\textheight}
{\includegraphics{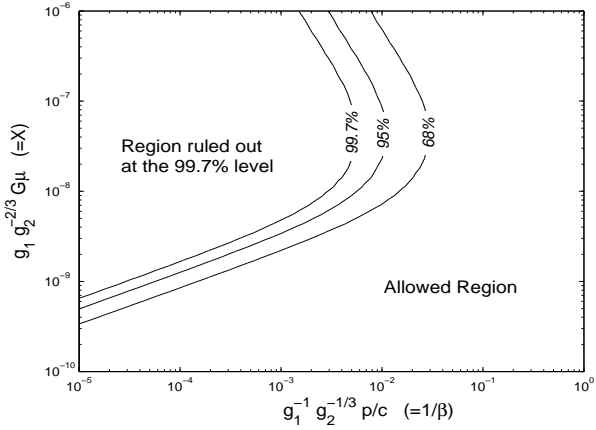}} \par}
\caption{Contour of excluded regions at various confidences. In the 
cosmic string model we have set $\tilde \varepsilon = 1$, and $n=3/2$,
and allowed $X$ and $\beta$ to vary. For the loudest event we have
used our Initial LIGO estimate for the amplitude $A_{50\%} =
10^{-20}$~s$^{-1/3}$.}
\label{fig:contours}
\end{figure}

If we write $\alpha = \varepsilon (\Gamma G \mu)^n$, and substitute
into Eqs.~(\ref{eq:theta3}) and (\ref{eq:zofA3}), we can simultaneously
absorb $g_1$ and $g_2$ into new variables $\tilde \varepsilon
(\varepsilon,g_1,g_2,n)$ and $X (G\mu, g_1, g_2)$. Looking at
Eq.~(\ref{eq:theta3}) we can write an identity for the ignorance
constants and the variables we want to absorb them into,
\begin{equation}
g_2 \varepsilon (\Gamma G \mu)^n = \tilde \varepsilon (\Gamma X)^n.
\label{eq:thetaabsorb1}
\end{equation}
Similarly, looking at the denominator of the right hand side of
Eq.~(\ref{eq:zofA3}) we write,
\begin{equation}
g_1 G\mu [\varepsilon (\Gamma G \mu)^n]^{2/3} 
= X [\tilde \varepsilon (\Gamma X)^n]^{2/3}
\label{eq:zofAabsrorb1}
\end{equation}
These equations can be simultaneously solved to give,
\begin{equation}
G\mu = g^{-1}_1 g_2^{2/3} X,
\label{eq:Gmuabsorb}
\end{equation}
and 
\begin{equation}
\varepsilon = g^{n}_1 g_2^{-2n/3-1} \tilde \varepsilon.
\label{eq:epsilonabsorb}
\end{equation}
If we replace $G\mu$ by Eq.~(\ref{eq:Gmuabsorb}) and $\varepsilon$ by
Eq.~(\ref{eq:epsilonabsorb}), in Eq.~(\ref{eq:theta3}) we obtain,
\begin{equation}
\theta_m 
= ((1+z)f_* \tilde \varepsilon (\Gamma X)^n H_0^{-1} \varphi_t
(z))^{-1/3},
\label{eq:theta4}
\end{equation}
and if we make the same replacements into Eq.~(\ref{eq:zofA3}), 
\begin{equation}
\frac{\varphi^{2/3}_t (z)}{(1+z)^{1/3} \varphi_r (z)} = 
\frac{A H^{-1/3}_0
}{X (\tilde \varepsilon (\Gamma X)^n)^{2/3}},
\label{eq:zofA4}
\end{equation}
namely, we obtain functions of $\tilde \varepsilon$ and $X$ only.

Finally, if we replace $G\mu$ by Eq.~(\ref{eq:Gmuabsorb}) and
$\varepsilon$ by Eq.~(\ref{eq:epsilonabsorb}), in our expression for the
rate, Eq.~(\ref{eq:rate12}), we see that we can absorb the remaining
factors by defining a quantity,
\begin{equation}
\beta = g_2^{1/3} g_1 \frac{c}{p}.
\label{eq:betaabsorb}
\end{equation}
This yields a rate,
\begin{eqnarray}
\frac {dR}{dz} 
&=& H_0 \frac{\beta(f_* H^{-1}_0)^{-2/3}}{2 
(\tilde \varepsilon (\Gamma X)^n)^{5/3} 
\Gamma X} \varphi^{-14/3}_t(z) \varphi_V(z) (1+z)^{-5/3}
\nonumber
\\
&\times& \Theta(1-\theta_m),
\label{eq:rate13}
\end{eqnarray}
where $\theta_m$ given by Eq.~(\ref{eq:theta4}). The parameters that
we constrain are now $X$, $n$, $\tilde \varepsilon$, and $\beta$.

Figure~\ref{fig:contours} illustrates the application of this
procedure. We show a contour plot constructed by computing the rate
and comparing it to $\gamma_{68\%} \approx 1.14/T$, $\gamma_{95\%}
\approx 3.00/T$ and $\gamma_{99.7\%} \approx 5.81/T$. In the cosmic
string model we have set $\tilde \varepsilon = 1$, and $n=3/2$, and
allowed $X$ and $\beta$ to vary. In place of the loudest event we have
used our Initial LIGO estimate for the amplitude $A_{50\%} =
10^{-20}$~s$^{-1/3}$. If we set the ignorance constants to unity, and
$c=1$, we see that for $G\mu$ in the range $4\times 10^{-9} -
10^{-6}$, values of the reconnection probability around $p = 10^{-3}$ are
ruled out at the $99.7\%$ level.

It is interesting to note that for cosmic superstrings a bound on the
reconnection probability might be turned into a bound on the string
coupling and/or a bound on the size of extra dimensions.

\section{Summary and Conclusions}

We begun this work by considering the data analysis infrastructure
necessary to perform a search for gravitational waves from cosmic
(super)string cusps. The optimal method to use in such a search is
matched-filtering. We have described the statistical properties of the
matched-filter output, a method for template bank construction, and an
efficient algorithm to compute the matched-filter. When data from
multiple interferometers is available, we discussed consistency checks
that can be used to greatly reduce the false alarm rate.

The relevant output of the matched-filter is the estimated signal
amplitude.  We have shown that in a search, the event with the largest
amplitude can be used to to set upper limits on the rate via the
loudest event method.  The upper limit depends on the cosmological
rate of events and the efficiency, which can be computed using
simulated signal injections into the data stream.  The injections
should account for certain properties of the population, such as the
frequency distribution ($dR/df \propto f^{-5/3}$ for cusps), and
source sky locations. We have briefly discussed the related issue of
sensitivity, which also depends on the search detection efficiency,
as well as background estimation. We have made a single estimate for
the amplitude of detectable and loudest events using the Initial LIGO
design curve and an SNR threshold $\rho_{\text{th}}=4$, which we believe is a
reasonable operating point for a pipeline. Assuming that there are no
significant inefficiencies in the pipeline, that the rate of false
alarms is sufficiently low, and that no signals or instrumental
glitches are present in the final trigger set, we expect this
amplitude to be $A_{50\%} \approx 10^{-20}$s$^{-1/3}$.  Advanced LIGO
is expected to be an order of magnitude more sensitive then Initial
LIGO, so for Advanced LIGO we expect $A_{50\%} \approx
10^{-21}$s$^{-1/3}$.

We have computed the rate of bursts in the case of an arbitrary cosmic
string loop distribution, using a generic cosmology. Motivated by the
data analysis, we cast the rate as a function of the amplitude. We
have applied this formalism in the case of a flat universe with
matter, radiation and a cosmological constant, to the loop
distribution used in the estimates of \cite{DV2,DV0,DV1}. For the
``classic'' (field theoretic strings with loops of size $l = \Gamma G
\mu t $) model of cosmic strings considered in \cite{DV0,DV1}, we find
substantially lower event rates than their estimates suggest.  The
bulk of the difference arises from our estimate of a detectable
amplitude. The effect of the amplitude estimate on the rate is
discussed in detail in Appendix B. The remaining discrepancy arises
from differences in the cosmology (the cosmological model in
\cite{DV2,DV0,DV1} included only matter and radiation), as well as
factors of $O(1)$ that were dropped in the previous estimates. The net
effect is that the chances of seeing an event from ``classic'' strings
using Initial LIGO data have dropped from order unity to about
$10^{-3}$ at the local maximum of the rate (which comes from bursts
produced in the matter era). This result is illustrated in
Fig.~\ref{fig:f1}.

The ``classic'' string model may well be incorrect so these results
are not necessarily discouraging. Two new results indicate that
this is the case. 

Recently it was realised that damping of perturbations propagating on
strings due to gravitational wave emission is not as efficient as
previously thought \cite{KandIGW}. As a consequence, the size of the
small-scale structure is sensitive the spectrum of perturbations
present on the strings, and can be much smaller than the canonical
value $\Gamma G \mu t$ \cite{trio}.  If the size of loops is given by
gravitational back-reaction, a reasonable guess for the spectrum of
perturbations on long strings leads to
\begin{equation}
\alpha \sim \left( \Gamma G \mu \right)^{3/2},
\label{eq:alpha3}
\end{equation}
which in turn could lead to detectable events by Advanced LIGO, or
Initial LIGO if we are dealing with cosmic superstrings. This is
illustrated in Figs.~\ref{fig:f2} and ~\ref{fig:f3}.

Even more recently, simulations \cite{RSB,shellardrecentsim,VOV}
suggest that loops are produced at sizes unrelated to the
gravitational back-reaction scale, $\alpha \lesssim 1$. In this case
loops of many different sizes are present at any given time (because
they are very long-lived), and the use of a revised loop distribution
becomes necessary. The particulars of the distribution are currently
under debate, but all distributions currently considered lead to an
enhanced rate of bursts relative to the ``classic'' model.  Results
for the rate for various loop distributions are shown in
Figs.~\ref{fig:f4} and \ref{fig:f5}.

Finally, we have have shown how the parameter space of theories that
lead to the production of cosmic strings might be constrained in the
absence of a detection. Results for a model of cosmic superstrings,
where the loop size is given by Eq.~(\ref{eq:alpha3}), are shown in
Fig.~\ref{fig:contours}. Initial LIGO may yield interesting
constraints on the reconnection probability for some range of string
tensions. Intriguingly, a bound on the reconnection probability might
be turned into a bound on the string coupling and/or a bound on the
size of extra dimensions, in the case of cosmic superstrings.

\subsection*{Wish list}

Unfortunately, there remain considerable uncertainties in models of
cosmic string evolution.  While we can estimate the detectability, and
place constraints, that are correct in the context of a particular
string model, the current parameter space allows for a wide spectrum
of burst rates in the interesting range of string tensions.

We would like to finish by posing a set of questions to the various
cosmic string simulation groups which, from the perspective of
gravitational wave detection, would greatly improve predictability.

They are:
\begin{itemize}
\item What is the size of cosmic string loops? 
Are they large when they are formed, so we need to consider a loop
distribution?  If so, what is that distribution? Is their size instead
given by gravitational back-reaction?  If so, what is the spectrum of
perturbations on long strings?
\item What is the number of cusps per loop oscillation? Is it 
independent of the loop size?
\item What is the size of cusps? In particular, what fraction of the 
loop length $l$ is involved in the cusp?
\item What are the effects of low reconnection probability? In 
particular, what is the enhancement in the loop density that results?
\end{itemize}

\begin{acknowledgements}
We are especially grateful to Alex Vilenkin and Ken Olum for carefully
reading the paper and suggesting several important improvements and
corrections. We are also grateful to Vicky Kalogera, Richard
O'Shaughnessy and Thibault Damour for carefully reading the paper and
providing many helpful suggestions and improvements. Finally, we would
like to thank Benjamin Wandelt and Marialessandra Papa for useful
discussions, Chris Ringeval for useful and friendly correspondence,
and Daniel Sigg for providing the analytic form for the LIGO design
curve, Eq.~(\ref{eq:LIGOIdesign}). The work of XS, JC, SRM, JR, and KC
was supported by NSF grants PHY 0200852 and PHY 0421416. The work of
IM was supported by the DOE and NASA.
\end{acknowledgements}

\appendix
\section{Cosmological Functions}

To derive exact expressions for $\varphi_t(z)$, $\varphi_r(z)$,
and $\varphi_V(z)$, we begin with the evolution of the Hubble
function. The Hubble function is given by
\begin{eqnarray}
 H(z)  =  H_0 h(z) \label{H},
\label{eq:H}
\end{eqnarray}
with, for a flat universe,
\begin{eqnarray}
h(z)  =  \left(\Omega_m(1+z)^3+\Omega_r(1+z)^4+
            \Omega_{\Lambda} \right)^{1/2}.
\label{eq:h}
\end{eqnarray}
Here, $\Omega_i=\rho_i(z=0)/\rho_c(z=0)$ is the present energy density
of the i'th component, relative to the critical density. The
subscripts $m$, $r$, and $\Lambda$ stand for matter (dark and
baryonic), radiation, and the cosmological constant,
respectively. Since we assume the universe flat, $\Sigma\Omega_i=1$.

We will use the set of cosmological parameters in \cite{stdcosmoref},
which provide a good fit to recent cosmological data.  The precise
values of the parameters are not critical, but we include them here
for clarity.  They are, $H_0=73$~km~s$^{-1}$Mpc$^{-1}=2.4
\times 10^{-18}$~s$^{-1}$, $\Omega_m=0.25$, $\Omega_r= 4.6 \times
10^{-5}$, and $\Omega_{\Lambda}=1-\Omega_m-\Omega_r$. 

To compute the relation between the time and the redshift, we use
the fact that $dz/dt=-(1+z)H$, and write
\begin{eqnarray}
 t & = & \int_0^t dt'=
      \int_z^{\infty}\frac{dz'}{(1+z')H(z')}.
\end{eqnarray}
The dimensionless function $\varphi_t(z)$ of Eq.~(\ref{eq:cosmict1}) is
thus,
\begin{eqnarray}
\varphi_t(z)  = \int_{z}^{\infty}\frac{dz'}{(1+z')h(z')}.
\label{eq:phitexact}
\end{eqnarray}

In order to compute the amplitude distance \cite{DV1} as a
function of the redshift, we consider null geodesics in an FRW
universe. We can take the polar and azimuthal coordinates to be
constant and let the radial coordinate, $r$, vary. In this case we
have that $dr/dt=-(1+z)$, and thus $dr/dz=1/H$. So we write
\begin{eqnarray}
r=\int_0^r dr'=\int_0^z\frac{dz'}{H(z')}. 
\label{eq:r}
\end{eqnarray}
We can therefore express the dimensionless function of Eq.~(\ref{eq:rofZ}),
$\varphi_r(z)$, as
\begin{eqnarray}
\varphi_r(z) & = & \int_0^{z}\frac{dz'}{h(z')}.
\label{eq:phiAexact}
\end{eqnarray}

Finally, we would like to derive an expression for the differential
volume as a function of the redshift, $dV/dz$. The differential volume
element is given by
\begin{eqnarray}
 dV & = & a^3(t)r^2dr \, \sin \theta d\theta \, d\phi ~,\nonumber
\end{eqnarray}
where $a(t)$ is the scale factor. Integrating over the polar and
azimuthal coordinates and using $dr/dz$ gives
\begin{eqnarray}
 dV  = 4\pi a^3(t)r^2 dr &=& \frac{4\pi r^2}{(1+z)^3H(z)}dz ~.
\end{eqnarray}
Using Eqs.~(\ref{eq:H}) and (\ref{eq:r}) gives the dimensionless 
function of Eq.~(\ref{eq:dv1})
\begin{eqnarray}
 \varphi_V(z) =  \frac{4\pi\varphi_r^2(z)}{(1+z)^3 h(z)}.
\label{eq:phiVexact}
\end{eqnarray}

\section{Approximate analytic expression for the rate}

In this appendix we compute an approximate expression for the rate as
a function of the amplitude can be obtained using the interpolating
functions introduced in \cite{DV2,DV0,DV1}, Eqs.~(\ref{eq:phit1}),
(\ref{eq:phiA1}), and (\ref{eq:phiV1}). The result is useful to understand
the qualitative behaviour of the rate curves shown in
Figs.~\ref{fig:f1}, \ref{fig:f2}, and \ref{fig:f3}.

Using Eqs.~(\ref{eq:amp11}) and (\ref{eq:phiA1}) we can write the
amplitude of an burst from a loop of length $l$ at a redshift $z$ as
\begin{equation}
A \sim \frac{G\mu l^{2/3} (1+z)^{2/3}}{t_0 z}.
\label{eq:amp111}
\end{equation}
We take the size of the feature that produces the cusp to be the
typical size of loops, $l(z) = \alpha H^{-1}_0 \varphi_t (z)$, with
$\varphi_t (z)$ given by Eq.~(\ref{eq:phit1}), so that,
\begin{equation}
A \sim G\mu \alpha^{2/3} t_0^{-1/3} z^{-1}(1+z)^{-1/3}(1+z/z_{eq})^{-1/3}.
\label{eq:amp12}
\end{equation}
We define a dimensionless amplitude $a$,
\begin{equation}
a \equiv \frac{A}{G\mu \alpha^{2/3} t_0^{-1/3}} 
\sim z^{-1}(1+z)^{-1/3}(1+z/z_{eq})^{-1/3}
\label{eq:a1}
\end{equation}
Depending on the value of $z$, the function $a(z)$ has three different
asymptotic behaviours,
\begin{equation}
a(z) \sim
\left\{
\begin{array}{ccc}
z^{-1}   & z \ll 1\\
z^{-4/3} & 1 \ll z \ll z_{eq}\\
z_{eq}^{1/3} z^{-5/3}&  z \gg z_{eq} 
\end{array}
\right.
\label{eq:a2}
\end{equation}
The regime where $z \ll 1$ corresponds to cusp events occurring nearby,
the regime where $1 \ll z \ll z_{eq}$ corresponds to matter-era events,
and the regime where $z \gg z_{eq}$ corresponds to radiation era
events.

At $z=1$, we define $$a_1 \equiv a(z=1) \approx 2^{-1/3} \sim 1,$$ and
at $z=z_{eq}$, $$a_{eq} \equiv a(z=z_{eq}) \approx z_{eq}^{-4/3}
2^{-1/3} \approx z_{eq}^{-4/3} a_1\approx 4\times 10^{-6}.$$ 
This means we can write,
\begin{equation}
z_{eq} \sim a_{eq}^{-3/4}.
\label{eq:zeqofaeq}
\end{equation}

The regime where $z \ll 1$,
corresponds to $a \gg 1$; the regime where $1 \ll z \ll z_{eq}$,
corresponds to $ a_{eq} \ll a \ll 1$; and the regime where $z \gg
z_{eq}$, corresponds to $a \ll a_{eq}$.  This means that we can write,
\begin{equation}
z(a) \sim
\left\{
\begin{array}{ccc}
a_{eq}^{-3/20} a^{-3/5}   & a \ll a_{eq}\\
a^{-3/4} & a_{eq} \ll a \ll 1\\
a^{-1}&  a \gg 1 
\end{array}
\right.
\label{eq:zofa}
\end{equation}
We can write this in terms of the interpolating function,
\begin{equation}
z(a) \sim a_{eq}^{-3/20} a^{-3/5} 
\left( 1+\frac{a}{a_{eq}}\right)^{-3/20}
\left( 1+ a\right)^{-1/4}.
\label{eq:zofa2}
\end{equation}

Using Eqs.~(\ref{eq:phit1}) and (\ref{eq:phiV1}), we can write the rate of
events as a function of the redshift, Eq.~(\ref{eq:rate12}), as,
\begin{equation}
\frac{dR}{dz} \sim b z^2 (1+z)^{-7/6} (1+z/z_{eq})^{11/6},
\label{eq:rate4}
\end{equation}
where $b$ is defined as
\begin{equation}
b \equiv 10^2 c \alpha^{-5/3} (p \Gamma G \mu)^{-1} t_0^{-1}
(f_* t_0)^{-2/3},
\label{eq:b1}
\end{equation}
and the theta-function cutoff has been ignored for convenience. In the
three regimes considered above the rate is given by,
\begin{equation}
dR \sim b \times
\left\{
\begin{array}{ccc}
z^{2}dz   & z \ll 1\\
z^{5/6}dz & 1 \ll z \ll z_{eq}\\
z_{eq}^{-11/6} z^{8/3}dz&  z \gg z_{eq} 
\end{array}
\right.
\label{eq:rate5}
\end{equation}

In the regime where  $z \ll 1$, corresponds to $a \gg 1$, and $z\sim a^{-1}$.
This means, 
\begin{equation}
dR \sim b z^2 dz = b a^{-2} \frac{dz}{da}da.
\label{eq:rate1stregime1}
\end{equation}
Since,
\begin{equation}
\frac{dz}{da} \sim -a^{-2},
\label{eq:rate1stregime2}
\end{equation}
we have that 
\begin{equation}
\begin{array}{ccc}
dR \sim -b a^{-4} da,  &\, \text{for} \,&  a \gg 1.\\
\end{array}
\label{eq:rate1stregime3}
\end{equation}

The regime where $1 \ll z \ll z_{eq}$, corresponds to $ a_{eq} \ll a
\ll 1$, and $z\sim a^{-3/4}$. This means,
\begin{equation}
dR \sim b z^{5/6} dz = b a^{-5/8} \frac{dz}{da}da.
\label{eq:rate2ndregime1}
\end{equation}
Since,
\begin{equation}
\frac{dz}{da} \sim -\frac{3}{4}a^{-7/4},
\label{eq:rate2ndregime2}
\end{equation}
we have that 
\begin{equation}
\begin{array}{ccc}
dR \sim -\frac{3}{4}ba^{-19/8}da,  &\, \text{for} \,&  a_{eq} \ll a \ll 1.\\
\end{array}
\label{eq:rate2ndregime3}
\end{equation}

Finally, the regime where $z \gg z_{eq}$, corresponds to $a \ll
a_{eq}$, and $z \sim (a_{eq}/a_1)^{-3/20} a^{-3/5}$. Thus, using
Eq.~(\ref{eq:zeqofaeq}),
\begin{equation}
dR \sim b z_{eq}^{-11/6} z^{8/3} dz = 
b a_{eq}^{39/40} a^{-8/5} 
\frac{dz}{da}da.
\label{eq:rate3rdregime1}
\end{equation}
Since,
\begin{equation}
\frac{dz}{da} \sim -\frac{3}{5} a_{eq}^{-3/20} a^{-8/5},
\label{eq:rate3rdregime2}
\end{equation}
we have that 
\begin{equation}
\begin{array}{ccc}
dR \sim -\frac{3}{5}b  a_{eq}^{33/40} 
a^{-16/5}da,  &\, \text{for} \,&  a \ll a_{eq}.\\
\end{array}
\label{eq:rate3rdregime3}
\end{equation}

Summarising,
\begin{equation}
\frac{dR(a)}{da} \sim
- b\times \left\{
\begin{array}{ccc}
a_{eq}^{33/40} a^{-16/5}   & a \ll a_{eq}\\
a^{-19/8} & a_{eq} \ll a \ll 1\\
a^{-4}&  a \gg 1 
\end{array}
\right.
\label{eq:Rofa1}
\end{equation}
Eq.~(\ref{eq:Rofa1}) can be integrated, to give the rate
of events with reduced amplitude greater than $a$,
\begin{equation}
R_{>a} \sim
b\times \left\{
\begin{array}{ccc}
a_{eq}^{33/40} a^{-11/5}   & a \ll a_{eq}\\
a^{-11/8} & a_{eq} \ll a \ll 1\\
a^{-3}&  a \gg 1 
\end{array}
\right.
\label{eq:Rgta1}
\end{equation}

We can determine the functional dependence of the rate for the
simple case when $\alpha = \Gamma G \mu$, which we can then compare
with the dashed and dashed-dot curves of Fig.~\ref{fig:f1}. The
factor of $b$, as defined by Eq.~(\ref{eq:b1}), contains a factor of
$(G\mu)^{-1}$ as well as a factor of $(G\mu)^{-5/3}$ through its
dependence on $\alpha$. So we take $b \propto (G\mu)^{-8/3}$.  The
dimensionless amplitude $a$, as defined by Eq.~(\ref{eq:a1}) also
contains a factor of $(G\mu)^{-1}$ as well as a factor of
$(G\mu)^{-2/3}$ through its dependence on $\alpha$, so that $a \propto
(G\mu)^{-5/3}$.

In Eq.~(\ref{eq:Rgta1}), (as we have mentioned) the regime where $a \ll
a_{eq}$ maps into the radiation era, and in this case the rate,
\begin{equation}
R \propto G\mu.
\label{eq:RgtaRAD}
\end{equation}
The regime where $a_{eq} \ll a \ll 1$ maps into the matter era when
the redshift $z \gg 1$, and the rate in this case is,
\begin{equation}
R \propto (G\mu)^{-31/24}.
\label{eq:RgtaMAT}
\end{equation}
The final regime when $a \gg 1$ corresponds to bursts that are coming
from close by ($z \ll 1$), and the rate,
\begin{equation}
R \propto (G\mu)^{7/3}.
\label{eq:RgtaCLOSE}
\end{equation}
So we immediately see that the first steep rise in the rate as a
function of $G\mu$ (from left to right) of the dashed and dashed-dot
curves of Fig.~\ref{fig:f1} corresponds to bursts that are
coming from small redshifts, i.e. from $z \ll 1$.  The slight decrease
in the rate comes from bursts produced at large redshifts, but still
in the matter era, i.e. $1 \ll z \ll z_{eq}$, and the final
increase in the rate comes from bursts produced in the radiation era,
$z \gg z_{eq}$.

Eq.~(\ref{eq:Rgta1}) also makes it easy to understand the lower burst
event rates we find relative to the previous estimates of Damour and
Vilenkin \cite{DV2,DV0,DV1} (see the the dashed and dashed-dot curves
of Fig.~\ref{fig:f1}).  They compared the strain produced by
cosmic string burst events at a rate of 1 per year to a noise induced
SNR $1$ event, given the Initial LIGO design noise curve (see, for
example, the dashed lines in Fig. 1 of \cite{DV2}). In our treatment,
the amplitude that corresponds to is $A \approx
10^{-21}$s$^{-1/3}$. To make our amplitude estimate we have chosen an
SNR threshold of $4$, which, as proposed in \cite{DV2}, is a
reasonable operating point for a data analysis pipeline, and have
included the effects of the antenna pattern of the instrument, which
averages to a factor of $\sqrt{5}$. The combined effect is to increase
the amplitude estimate by about a factor of $10$, to make it $A\approx
10^{-20}$s$^{-1/3}$.

Looking at Eq.~(\ref{eq:Rgta1}), we see that an increase in the amplitude
of a factor of $10$, results in a decrease of a factor of $10^3$ in
the rate of nearby bursts ($z \ll 1$), a decrease by a factor of about
$24$ in the rate of bursts produced in the matter era ($1 \ll z \ll
z_{eq}$), and a decrease by about a factor of $160$ in the rate of
bursts produced in the radiation era ($z \gg z_{eq}$).

\end{document}